\documentclass[11pt, a4paper]{article}
\usepackage{jheparxiv}
\usepackage[utf8]{inputenc}
\usepackage{amsmath}
\usepackage{amsfonts}
\usepackage{amssymb}
\usepackage{latexsym}
\usepackage{mathrsfs}
\usepackage{braket}		
\usepackage{graphicx}
\usepackage{color}
\usepackage{xcolor}
\usepackage{slashed}
\usepackage{twistor}
\usepackage[all]{xy}
\usepackage{tikz-cd}
\usepackage{mathtools}

\newcommand{\veps}{\varepsilon}

\newcommand{\msf}[1]{\mathsf{#1}}

\renewcommand{\cO}{\mathscr{O}}

\title{All-order celestial OPE in the MHV sector}

\author[a]{Tim Adamo,}
\author[a]{Wei Bu,}
\author[b]{Eduardo Casali,}
\author[b,c]{Atul Sharma}

\affiliation[a]{School of Mathematics and Maxwell Institute for Mathematical Sciences\\
University of Edinburgh, EH9 3FD, UK}
\affiliation[b]{Center for the Fundamental Laws of Nature\\ Harvard University, Cambridge, MA, 02138, USA}
\affiliation[c]{Black Hole Initiative\\
Harvard University, Cambridge, MA, 02138, USA}

\emailAdd{t.adamo@ed.ac.uk}
\emailAdd{w.bu@sms.ed.ac.uk}
\emailAdd{ecasali@g.harvard.edu} 
\emailAdd{atulsharma@fas.harvard.edu}

\abstract{On-shell kinematics for gluon scattering can be parametrized with points on the celestial sphere; in the limit where these points collide, it is known that tree-level gluon scattering amplitudes exhibit an operator product expansion (OPE)-like structure. While it is possible to obtain singular contributions to this celestial OPE, getting regular contributions from both holomorphic and anti-holomorphic sectors is more difficult. In this paper, we use twistor string theory to describe the maximal helicity violating (MHV) sector of tree-level, four-dimensional gluon scattering as an effective 2d conformal field theory on the celestial sphere. By organizing the OPE between vertex operators in this theory in terms of soft gluon descendants, we obtain all-order expressions for the celestial OPE which include all regular contributions in the collinear expansion. This gives new, all-order formulae for the collinear splitting function (in momentum space) and celestial OPE coefficients (in the conformal primary basis) of tree-level MHV gluon scattering. We obtain these results for both positive and negative helicity gluons, and for any incoming/outgoing kinematic configuration within the MHV sector.}

\begin{document} 
\maketitle
\flushbottom

\section{Introduction}

It is a simple kinematical fact that massless 4-momenta can be parametrized by a point $(z,\bar{z})$ on the celestial sphere, along with a frequency $\omega$. In this parametrization, collinear limits of massless particles are described by collision of the corresponding points on the celestial sphere. In gauge theory or gravity, the amplitudes factorize in terms of well-known collinear splitting functions, whose leading behaviour in the collinear limit is proportional to a simple pole in the inner product between the collinear momenta (cf., \cite{Altarelli:1977zs,Mangano:1990by,Bern:1998sv}). All subleading corrections to the splitting function are regular in the inner product of collinear momenta, and the factor multiplying the splitting function is itself a lower-point scattering amplitude. 

On the celestial sphere, the structure of the collinear limit resembles, at least heuristically, an operator product expansion (OPE). It turns out that this can be made precise: replacing the frequency $\omega$ of each external particle with a conformal dimension $\Delta$ by means of a Mellin transform turns the scattering amplitude into an object which transforms like a conformal correlation function on the celestial sphere~\cite{Pasterski:2016qvg,Pasterski:2017kqt}. The resulting Mellin-transformed object is often called a celestial amplitude, and the external states are conformal primary wavefunctions. The Mellin transform precisely relates the collinear limit in momentum space with the OPE limit of colliding operator insertions on the celestial sphere~\cite{Fan:2019emx,Pate:2019lpp}. So with a slight abuse of terminology, one can refer to both the collinear limit in momentum space and the OPE limit in the conformal primary basis as \emph{celestial OPEs} defined on the sphere.

Celestial OPEs play a key role in celestial holography, which aims to describe massless S-matrix elements for `bulk' theories in flat spacetime in terms of a two-dimensional conformal field theory (CFT) (see~\cite{Pasterski:2021raf,McLoughlin:2022ljp} and references therein). Indeed, celestial OPEs encode the data (the OPE coefficients and spectrum) of such a celestial CFT (CCFT), so determining them is important for any bottom-up construction of celestial holography. The singular terms contributing to tree-level celestial OPEs have been determined by direct Mellin transform or symmetry arguments for a broad array of massless field theories~\cite{Fan:2019emx,Pate:2019lpp,Himwich:2021dau,Jiang:2021ovh}, and this in turn led to the discovery of various infinite-dimensional asymptotic symmetry algebras underlying gluon and graviton scattering~\cite{Guevara:2021abz,Strominger:2021lvk,Himwich:2021dau}. There have now been an extensive number of studies investigating the structural properties (e.g., associativity) of the singular contributions to tree- and loop-level celestial OPEs~\cite{Fan:2021isc,Guevara:2021tvr,Atanasov:2021cje,Fan:2021pbp,Costello:2022upu,Ren:2022sws,Monteiro:2022lwm,Bhardwaj:2022anh,Ball:2022bgg,Bittleston:2022jeq}. 

This singular OPE data would be enough to uniquely determine a local, unitary CFT, but it is clear that any CCFT will have exotic properties (e.g., distributional terms in 2- and 3-point correlation functions~\cite{Pasterski:2017ylz} and complicated analytic dependence on conformal dimensions~\cite{Schreiber:2017jsr}), and the singular celestial OPEs are not sufficient. To constrain CCFTs, one requires not just the singular terms, but the regular, subleading terms of the celestial OPE as well. Even in momentum space (where such terms represent subleading corrections to the collinear splitting function) these terms probe the degree to which four-dimensional scattering amplitudes are governed by an associative OPE of a two-dimensional CFT (celestial or otherwise). 

Starting with the work of~\cite{Banerjee:2020kaa}, there has been an effort to determine regular contributions to the celestial OPE using symmetry constraints or brute force calculations in the collinear expansion of low-multiplicity gluon or graviton amplitudes~\cite{Banerjee:2020zlg,Ebert:2020nqf,Banerjee:2020vnt,Banerjee:2021cly,Banerjee:2021dlm}. These strategies get complicated very quickly, and it is not at all clear how to proceed to arbitrary order in the OPE, even for specific scattering amplitudes at tree-level. One approach to tackling this problem is to find a \emph{dynamical} principle for computing celestial OPEs directly, rather than relying on kinematical constraints or Mellin transforming explicit momentum space calculations.

\medskip

In this paper, we compute \emph{all} of the regular terms in the tree-level celestial OPE between gluons for the maximal helicity violating (MHV) sector using \emph{twistor string theory}~\cite{Witten:2003nn,Berkovits:2004hg}. This builds on the deep and long-standing connections between twistor theory and the asymptotic geometry of spacetimes~\cite{Newman:1976gc,Penrose:1976js,Hansen:1978jz,Eastwood:1982,Adamo:2014yya,Geyer:2014lca,Adamo:2015fwa}. The twistor string provides a dynamical principle for direct calculation of the celestial OPE, and the connection between twistor theory and celestial holography has now been established in a variety of ways~\cite{Adamo:2021lrv,Costello:2021bah,Adamo:2021zpw,Bu:2021avc,Costello:2022wso,Costello:2022jpg}. In a prior paper~\cite{Adamo:2021zpw}, we showed that the worldsheet OPE of (ambi)twistor string theory localizes to the singular OPE limit on the celestial sphere and hence dynamically generates all singular terms in the celestial OPE without any truncation or approximations. However, it was not clear how to obtain regular terms in the OPE with this framework, which localized on the region of the string moduli space corresponding to the singular terms only.

By restricting our attention to the MHV sector, where the twistor string worldsheet is holomorphically identified with the celestial sphere, we obtain an effective description of tree-level MHV gluon scattering in terms of a two-dimensional CFT on the celestial sphere itself. This means that all terms in the celestial OPE (singular and regular) can be read off from the OPEs between vertex operators in this effective CFT. It is crucial that the resulting OPEs are organized in terms of soft gluon descendants to obtain expressions which close on the gluon vertex operators at each order in the OPE expansion. This provides an interesting contrast with the work of~\cite{Banerjee:2020zlg,Ebert:2020nqf,Banerjee:2020vnt,Banerjee:2021cly,Banerjee:2021dlm}, where the expansion is organized in terms of Kac-Moody and Virasoro descendants.  
 
Our results depend only on the fact that the two gluons involved in the celestial OPE are external states in a tree-level MHV amplitude; the multiplicity of this amplitude or the specific structure of the Parke-Taylor formula~\cite{Parke:1986gb} or its Mellin transform is irrelevant. This provides both an all-multiplicity proof of the subleading celestial OPE and an explicit realization of the OPE by a two-dimensional CFT on the sphere. The latter guarantees that the resulting MHV celestial OPE is associative as well as invariant under holographic symmetry algebras.

The all-order MHV gluon celestial OPEs are surprisingly compact. For instance, the OPE for positive helicity gluons with on-shell 4-momenta $k_{i}^{\alpha\dot\alpha}=\kappa_{i}^{\alpha}\,\tilde{\kappa}_i^{\dot\alpha}$ represented on the celestial sphere by vertex operators $\cU_{+}^{\msf{a}}(\kappa_i,\tilde{\kappa}_j)$, the celestial OPE for the MHV sector is:
\be
\cU^\msf{a}_{+}(\kappa_i,\tilde{\kappa}_i)\,\cU^\msf{b}_{+}(\kappa_j,\tilde{\kappa}_j) = \sum_{p=0}^{\infty}\la i\,j\ra^{p-1}\sum_{k=0}^p\sum_{\ell=0}^{p-k}\frac{(-[i\,\p_j])^{\ell}}{\ell!}\,J^\msf{a}_{-p}[k](\tilde\kappa_i)\,\cU_+^\msf{b}(\kappa_j,\tilde\kappa_i+\tilde\kappa_j)\,.
\ee
Here, $\msf{a},\,\msf{b}$ are the colour indices of the two gluons, $\la i\,j\ra=\kappa_i^{\alpha}\,\kappa_{j\,\alpha}$, $[i\,\partial_j]=\tilde{\kappa}_i^{\dot\alpha}\frac{\p}{\p\tilde{\kappa}_j^{\dot\alpha}}$, and $J^{\msf{a}}_{-p}[k]$ are the soft gluon descendants. This expression can easily be Mellin transformed to yield the all-order, tree-level celestial OPE in CCFT between positive helicity gluons in the MHV sector. For example, the result when both gluons are outgoing is
\begin{multline}
\cU_{+,\Delta_i}^{\msf{a}}(z_i,\bar{z}_i)\,\cU_{+,\Delta_j}^{\msf{b}}(z_j,\bar{z}_j)= \sum_{p=0}^{\infty}z_{ij}^{p-1}\sum_{k=0}^p\sum_{\ell=0}^{p-k}\sum_{m=0}^{\infty}\frac{\bar z_{ij}^m}{m!}\,\frac{\bar{D}_k^{\ell}}{\ell !}\, J^{\msf{a}}_{-p}[k](\bar z_i)\\
    \times\,B(\Delta_i+k+\ell+m-1,\,\Delta_j-1)\,\bar\partial^m_j\,\cU_{+,\Delta_i+\Delta_j+k-1}^{\msf{b}}(z_j,\bar z_j)\,,
\end{multline}
where $\Delta_i,\Delta_j$ are the conformal dimensions of the conformal primary wavefunctions, $z_{ij}=z_i-z_j$, $\bar{z}_{ij}=\bar{z}_i-\bar{z}_j$ denote the holomorphic and anti-holomorphic separations on the celestial sphere, $\bar{D}_k=-\bar{z}_{ij}\dbar_j+\Delta_i+\Delta_j+k-3$ is a differential operator on the celestial sphere, and $B(x,y)$ is the Euler Beta function. 

More generally, we obtain expressions for the tree-level celestial OPE between two gluons of positive or mixed positive/negative helicity in any incoming/outgoing configuration in the MHV sector. The intrinsic chirality of the MHV sector and twistor string theory means that there are no 
nontrivial OPE expressions for two negative helicity gluons that can be obtained in this framework.

\medskip

The paper is structured as follows: Section~\ref{Sec:TS} reviews the salient aspects of twistor string theory and the induced description of the tree-level MHV sector in terms of a 2d CFT on the celestial sphere. Section~\ref{Sec:OPE} then sets out the details of the celestial OPE computation, defining the soft gluon descendants used to organize the expansion and performing the calculation for both momentum eigenstates and the conformal primary basis. Section~\ref{Sec:Conc} concludes with a discussion of open problems and future directions. Appendix~\ref{app:exp} proves some identities used during the OPE computation.


\section{Twistor strings and the MHV sector on the celestial sphere}\label{Sec:TS}

The maximal helicity violating (MHV) sector of tree-level gluon scattering amplitudes -- with two negative helicity and arbitrarily many positive helicity external gluons -- is remarkably simple: the amplitudes are captured at all multiplicities by the famous Parke-Taylor formula~\cite{Parke:1986gb}. Nair observed long ago that the Parke-Taylor formula can be understood as a worldsheet correlator in twistor space~\cite{Nair:1988bq}, and Witten refined and generalized this to formulate tree-level Yang-Mills theory as a twistor string~\cite{Witten:2003nn}. 

In twistor string theory, the MHV sector corresponds to holomorphic maps from a Riemann sphere to twistor space which are \emph{linear}. When vertex operators for the external gluons correspond to momentum eigenstates or conformal primary wavefunctions -- parametrized by points on the celestial sphere and a frequency or scaling dimension, respectively -- this means that the worldsheet is holomorphically identified with the celestial sphere.

\medskip

Broadly speaking, twistor strings are 2d chiral worldsheet CFTs governing holomorphic maps from a closed Riemann surface to \emph{twistor space}, a complex projective space which is related non-locally to spacetime (see Section 3 of~\cite{Geyer:2022cey} for a recent review). Twistor strings have been formulated to describe several massless field theories, including supergravity~\cite{Skinner:2013xp} and ABJM theory~\cite{Engelund:2014sqa}, but we will be interested in the Berkovits-Witten twistor string describing four-dimensional Yang-Mills theory~\cite{Witten:2003nn,Berkovits:2004hg,Berkovits:2004jj,Reid-Edwards:2012vwx}. 

A fact about this twistor string theory is that genus zero correlation functions of vertex operators representing positive and negative helicity gluons are equal to tree-level gluon scattering amplitudes. The worldsheet path integral imposes a relationship between the degree of the holomorphic map from the worldsheet to twistor space and the number of negative helicity gluon vertex operators: maps of degree $d$ correspond to correlators involving $d+1$ negative helicity gluons. In other words, the helicity grading of the worldsheet correlator localizes the degree of the holomorphic map. After this localization, the resulting expression, often referred to as the Roiban-Spradlin-Volovich-Witten (RSVW) formula~\cite{Witten:2003nn,Roiban:2004yf}, gives the tree-level N$^{d-1}$MHV gluon scattering amplitude in Yang-Mills theory. The formula can be proved to correctly capture these scattering amplitudes using on-shell recursion relations~\cite{Skinner:2010cz,Dolan:2011za} or worldsheet factorization arguments~\cite{Adamo:2013tca}.

For our purposes, it suffices to simply state the gluon vertex operators and the OPEs between the worldsheet fields which result from quantizing the model at genus zero. Let $Z^{A}=(\mu^{\dot\alpha},\lambda_{\alpha})$ be homogeneous coordinates on projective twistor space, $\PT$, which is the open subset of $\CP^3$ corresponding to $\lambda_{\alpha}\neq 0$. This means that the $Z^A$ are considered only up to overall projective rescalings, $Z^A\sim r\,Z^A$ for any non-vanishing complex number $r\in\C^*$. As worldsheet fields on the Riemann sphere $\CP^1$, these are bosons with zero conformal weight, taking values in degree $d$ holomorphic maps $\CP^1\to\PT$.

Positive and negative helicity gluons are represented by the (integrated) vertex operators
\be\label{VOs}
\cU_+^{\msf{a}} = \int_{\Sigma} j^{\msf{a}}\, a(Z)\,, \qquad \cU_-^{\msf{a}} = \int_{\Sigma} \,j^{\msf{a}}\, \msf{O}\, b(Z)\,.
\ee
Here, $a$ and $b$ are twistor representatives for the positive and negative helicity gluons, respectively; they are valued in
\be\label{cohreps}
a\in H^{0,1}(\PT,\cO)\,, \qquad b\in H^{0,1}(\PT,\cO(-4))\,,
\ee
where $H^{0,1}(\PT,\cO(k))$ denotes the cohomology group on $\PT$ of $(0,1)$-forms which are homogeneous of degree $k$ and $\dbar$-closed but not exact. The Penrose transform, relating cohomology on $\PT$ to massless free fields in (complexified) Minkowski spacetime, ensures that these representatives correspond to on-shell positive and negative helicity gluons~\cite{Penrose:1969ae,Ferber:1977qx,Eastwood:1981jy}. In the negative helicity vertex operator, $\msf{O}$ is constructed from fermionic fields $\chi^{a}$, $a=1,\ldots,4$:
\be\label{fermO}
\msf{O}:=\frac{\epsilon_{abcd}}{4!}\,\chi^{a}\,\chi^{b}\,\chi^{c}\,\chi^{d}=\frac{\chi^{4}}{4!}\,,
\ee
where $\chi^a$ have vanishing conformal weight, are valued in degree $d$ holomorphic polynomials on the worldsheet and have the same projective scaling as the twistor coordinates $Z^A$. This ensures that the integrand of the negative helicity vertex operator is homogeneous of weight zero, and thus well-defined. 

The conformal weight $(1,0)$ currents $j^{\msf{a}}$ appearing in both vertex operators arise from a worldsheet current algebra, with the index $\msf{a}$ in the associated Lie algebra. The current algebra contributes all of the non-trivial Wick contractions in any correlation function of the vertex operators \eqref{VOs} through the OPE between the currents:
\begin{equation}
    j^{\msf{a}}(\sigma)\,j^{\msf{b}}(\sigma')\sim \frac{k\,\delta^{\msf{a}\msf{b}}}{(\sigma-\sigma')^2}+ \frac{f^{\msf{a}\msf{b}\msf{c}}\,j^{\msf{c}}(\sigma')}{\sigma-\sigma'}\,,
\end{equation}
where $\sigma,\sigma'$ are affine coordinates on the genus zero worldsheet, $k$ is the level of the current algebra and $f^{\msf{abc}}$ are the structure constants of the associated gauge group. The double pole in this OPE leads to gravitationally-mediated double-trace terms in correlation functions, so to decouple these gravitational degrees of freedom (which correspond to four-derivative conformal gravity, so are also non-unitary) we set the level of the worldsheet current algebra to zero: $k\to0$~\cite{Berkovits:2004jj,Adamo:2018hzd}\footnote{Ordinarily the level is required to be a positive integer for the worldsheet current algebra to be globally well-defined, but for calculations based only on local worldsheet OPEs, having $k=0$ is not problematic.}. Hence, the current OPE on the worldsheet will be taken to have only a simple pole
\begin{equation}\label{jj_OPE}
    j^{\msf{a}}(\sigma)\,j^{\msf{b}}(\sigma')\sim  \frac{f^{\msf{a}\msf{b}\msf{c}}\,j^{\msf{c}}(\sigma')}{\sigma-\sigma'}\,,
\end{equation}
from now on.

To obtain answers in momentum space, one chooses momentum eigenstate representatives for the twistor representatives in the vertex operators \eqref{VOs}. For an on-shell 4-momentum decomposed into spinors as $k^{\alpha\dot\alpha}=\kappa^{\alpha}\tilde{\kappa}^{\dot\alpha}$, the corresponding positive and negative helicity gluon representatives in twistor space are (cf., \cite{Adamo:2011pv}):
\be\label{momeig0}
    a(Z)=\int_{\C^*}\frac{\d s}{s}\,\bar{\delta}^{2}(\kappa-s\,\lambda)\,\e^{\im\,s[\mu\,\tilde{\kappa}]}\,, \qquad b(Z)=\int_{\C^*}\d s\,s^{3}\,\bar{\delta}^{2}(\kappa-s\,\lambda)\,\e^{\im\,s[\mu\,\tilde{\kappa}]}\,,
\ee
with the scale parameter $s\in\C^{*}$ ensuring that each representative has the appropriate homogeneity on twistor space and
\be\label{holdelt}
\bar{\delta}^{2}(\kappa-s\,\lambda):=\frac{1}{(2\pi\im)^2}\bigwedge_{\alpha=0,1}\dbar\left(\frac{1}{\kappa_{\alpha}-s\,\lambda_{\alpha}}\right)\,,
\ee
is the two-dimensional holomorphic delta function.

\medskip

Now, given the relation between the degree $d$ of the holomorphic map from the worldsheet to twistor space, and the helicity grading of the resulting scattering amplitude (namely, N$^{d-1}$MHV), the MHV sector of twistor string theory corresponds to linear, $d=1$ holomorphic maps. In this case, the zero modes of the $Z^{A}$ worldsheet fields are 
\be\label{wszm1}
Z^{A}(\sigma)=U^{A}_{\alpha}\,\sigma^{\alpha}\,,
\ee
where $\sigma_{\alpha}=(\sigma_0,\sigma_1)$ are homogeneous coordinates on the $\CP^1$ worldsheet and $\{U^{A}_{\alpha}\}$ are the map moduli. The remaining GL$(2,\C)$ redundancy in this description can be fixed by picking (compatible) values for four of the moduli; a particularly convenient choice is to set $U^{\beta}_{\alpha}=\delta^{\alpha}_{\beta}$. This holomorphically identifies the homogeneous coordinates $\sigma_{\alpha}$ on the worldsheet $\Sigma\cong\CP^1$ with the $\lambda_{\alpha}$ components of the twistor field. With this identification, the remaining components of the degree one map to twistor space are simply
\begin{equation}\label{worldsheet_incidence}
\mu^{\dot\alpha}(\lambda)=x^{\alpha\dot\alpha}\,\lambda_{\alpha}\,, \qquad \chi^{a}(\lambda)=\theta^{a\alpha}\,\lambda_{\alpha}\,,
\end{equation}
where $(x^{\alpha\dot\alpha},\theta^{a\alpha})$ are the remaining four bosonic and eight fermionic moduli. 

The relations \eqref{worldsheet_incidence} are easily recognizable as the standard twistor incidence relations for chiral Minkowski superspace, relating a point $(x,\theta)\in\M$ to a holomorphic, linearly embedded sphere in $\PT$; for the MHV sector, this Riemann sphere in twistor space is identified (holomorphically) with the string worldsheet itself. Furthermore, the sphere in twistor space with homogeneous coordinates $\lambda_{\alpha}$ is precisely the celestial sphere (cf., \cite{Adamo:2021lrv}).

This holomorphic identification between the twistor string worldsheet and the celestial sphere for the MHV sector can be implemented directly at the level of the individual vertex operators. This allows us to effectively describe the MHV sector in terms of a 2d CFT living on the celestial sphere itself. To do this, we parametrize the on-shell 4-momentum of the $i^{\mathrm{th}}$ external gluon as $k_i^{\alpha\dot\alpha}=\kappa_i^{\alpha}\,\tilde{\kappa}_{i}^{\dot\alpha}$ with
\be\label{momparam}
\kappa_{i\,\alpha}=(1,\,z_i)\,, \qquad \tilde{\kappa}_{i\,\dot\alpha}=\varepsilon_i\,\omega_i\,\bar{z}_{i\,\dot\alpha}=\varepsilon_i\,\omega_i\,(1,\,\bar{z}_i)\,,
\ee
where $\omega_i$ is the frequency, $(z_i,\bar{z}_i)$ is a point on the celestial sphere (in an affine coordinate patch), and $\varepsilon_i=\pm1$ denotes whether the momentum in incoming ($\varepsilon_i=-1$) or outgoing ($\varepsilon_i=1$). Note that this is a slightly non-standard parametrization, associating the frequency with the anti-holomorphic coordinates on the celestial sphere. This simplifies subsequent formulae by identifying holomorphic momentum spinor contractions with holomorphic displacements on the celestial sphere, $\la i\,j\ra=z_i-z_j$, the only price is that holomorphic coordinates on the celestial sphere now carry little group weight. 

Consider a positive helicity gluon vertex operator $\cU_{+,i}^{\msf{a}}$ in the MHV sector of the twistor string. With the GL$(2,\C)$-fixing of \eqref{worldsheet_incidence}, for which $\sigma_{i\,\alpha}=\lambda_{i\,\alpha}$, we choose an affine patch of the Riemann sphere so that $\lambda_{i\,\alpha}=(1,\lambda_i)$. The holomorphic delta function in the vertex operator can be decomposed using a normalized spinor dyad $a^{\alpha}=(1,0)$, $b^{\alpha}=(0,-1)$:
\be\label{EFTVO1}
\begin{split}
\cU^{\msf{a}}_{+,i}&=\int\limits_{\CP^1\times\C^*}j^{\msf{a}}(\lambda_i)\,\frac{\d s_i}{s_i}\,\bar{\delta}(\la a\,i\ra-s_i\,\la a\,\lambda_i\ra)\,\bar{\delta}(\la b\,i\ra-s_i\,\la b\,\lambda_i\ra)\,\e^{\im\,s_i\,[\mu(\lambda_i)\,i]} \\
 &=\int\limits_{\CP^1\times\C^*}j^{\msf{a}}(\lambda_i)\,\frac{\d s_i}{s_i}\,\bar{\delta}(1-s_i)\,\bar{\delta}(s_i\,\lambda_i-\,z_i)\,\e^{\im\,s_i\,[\mu(\lambda_i)\,i]}\,,
\end{split}
\ee
with $\la a\,b\ra=1$. The scale integral in $s_i$ can now be performed against the first delta function, leaving
\be\label{EFTVO2}
\cU^{\msf{a}}_{+,i}=\int_{\CP^1}j^{\msf{a}}(\lambda_i)\,\bar{\delta}(\lambda_i-z_i)\,\e^{\im\,[\mu(\lambda_i)\,i]}\,,    
\ee
with the remaining delta function explicitly implementing the holomorphic identification between the vertex operator insertion on the worldsheet ($\lambda_i$) and the point on the celestial sphere corresponding to the gluon momentum ($z_i$). 

The worldsheet integral in \eqref{EFTVO2} can now be done against this remaining delta function to leave a vertex operator defined entirely on the celestial sphere:
\be\label{effective_vertex_operator_+}
\cU^{\msf{a}}_{+,i}(z_i,\tilde{\kappa}_i) = j^\msf{a}(z_i)\,\exp\left(\im\,[\mu(z_i)\,i]\right)\,.
\ee
Here, we have abused notation by denoting the current $j^{\msf{a}}$ with the same symbol that previously denoted the worldsheet current of conformal weight $(1,0)$; in particular $j^{\msf{a}}_{\text{old}}(\sigma)=j^{\msf{a}}_{\text{new}}(\sigma)\,\d\sigma$ in terms of affine coordinates. From now on, $j^{\msf{a}}$ denotes the Kac-Moody current with zero conformal weight. An identical procedure can be used to reduce the negative helicity gluon vertex operator to a vertex operator on the celestial sphere:
\be\label{effective_vertex_operator_-}
\cU^{\msf{a}}_{-,i}(z_i,\tilde{\kappa}_i) = j^\msf{a}(z_i)\,\msf{O}(z_i)\,\exp\left(\im\,[\mu(z_i)\,i]\right)\,,
\ee
with $j^{\msf{a}}$ again denoting the pure conformal weight zero Kac-Moody current.

It is straightforward to Mellin transform these expressions to obtain the corresponding celestial sphere vertex operators in the conformal primary basis~\cite{Pasterski:2016qvg,Pasterski:2017kqt}. This Mellin transform exchanges the frequency $\omega_i$ with a scaling dimension $\Delta_i$:
\be\label{cVOs}
\cU^{\msf{a},\varepsilon_i}_{\pm,\Delta_i}(z_i,\bar{z}_i)=\int_{0}^{\infty}\d\omega_{i}\,\omega_i^{\Delta_i-1\mp1}\,\cU^{\msf{a}}_{\pm,i}(z_i,\tilde{\kappa}_i)\,,
\ee
where the additional helicity dependent factors of $\omega_i$ arise due to the non-standard little group scaling associated with the parametrization \eqref{momparam}. For these conformal primary vertex operators, the incoming/outgoing parameter is explicit, as the Mellin transform strips off the prefactors $\varepsilon_i\,\omega_i$ of the momentum spinor $\tilde{\kappa}_i$.

\medskip

Regardless of whether one considers momentum eigenstates or conformal primary wavefunctions, it is clear that the positive and negative helicity gluon vertex operators can be reduced to operator insertions on the celestial sphere for the MHV sector. Interactions between these vertex operators can now be captured by an effective 2d CFT defined on the celestial sphere itself:
\be\label{2dEFT}
S^{\text{MHV}}=\frac{1}{2\pi}\int_{\CP^1}\tilde{\lambda}_{\dot\alpha}(z)\,\dbar\mu^{\dot\alpha}(z)+S_{\text{current}}\Big|_{\CP^1}\,,
\ee
where $S_{\text{current}}|_{\CP^1}$ denotes the worldsheet current algebra restricted to the celestial sphere. This CFT induces OPEs between the fields on the celestial sphere:
\be\label{CSOPE}
\mu^{\dot\alpha}(z_i)\,\tilde{\lambda}_{\dot\beta}(z_j)\sim\frac{\delta^{\dot\alpha}_{\dot\beta}}{z_{ij}}\,, \qquad j^{\msf{a}}(z_i)\,j^{\msf{b}}(z_j)\sim\frac{f^{\msf{abc}}\,j^{\msf{c}}(z_j)}{z_{ij}}\,,
\ee
where $z_{ij}:=z_i-z_j$.

This 2d CFT is `effective' in the sense that it only describes correlation functions of the vertex operators in the MHV configuration. Indeed, it is easy to see that these correlators reproduce the Parke-Taylor formula for the MHV amplitude. Taking only one colour-ordered term contributing to the correlation function of Kac-Moody currents, one finds
\begin{multline}\label{EFTcorr}
\left\la \cU_{-,i}\,\cU_{-,j}\,\prod_{l\neq i,j}\cU_{+,l}\right\ra=\frac{1}{z_{12}\,z_{23}\cdots z_{n1}}\,\int\d^{4}x\,\d^{4}\theta\,\msf{O}(z_i)\,\msf{O}(z_j)\,\exp\left(\im\sum_{l=1}^{n}k_l\cdot x\right) \\
 =(2\pi)^{4}\,\delta^{4}\!\left(\sum_{l=1}^{n}k_l\right)\,\frac{z_{ij}^4}{z_{12}\,z_{23}\cdots z_{n1}} = (2\pi)^{4}\,\delta^{4}\!\left(\sum_{l=1}^{n}k_l\right)\,\frac{\la i\,j\ra^{4}}{\la1\,2\ra\,\la2\,3\ra\cdots\la n\,1\ra}\,,
\end{multline}
as desired, having used
\be\label{Grassint}
\int \d^{8}\theta\,\msf{O}(z_i)\,\msf{O}(z_j)=z_{ij}^{4}\,,
\ee
when evaluating the fermionic moduli integral. 

More generally, OPEs between fields or the vertex operators \eqref{effective_vertex_operator_+} \eqref{effective_vertex_operator_-}, or \eqref{cVOs} defined by the CFT \eqref{2dEFT} only hold in the MHV sector. Within this sector, such OPEs in the 2d CFT are, by definition, OPEs on the celestial sphere itself.


\section{OPE computation}\label{Sec:OPE}

Armed with the description of the MHV sector on the celestial sphere in terms of the vertex operators \eqref{effective_vertex_operator_+}, \eqref{effective_vertex_operator_-} and the 2d CFT \eqref{2dEFT}, we can now consider the OPE between gluon insertions. For momentum eigenstates, the resulting OPE expansion corresponds to an expansion in the holomorphic collinear limit, and by Mellin transforming the result we obtain the descendant expansion of the celestial OPE for the MHV sector of CCFT. 

Since the effective theory lives on the celestial sphere itself, we are able to perform the descendant expansions in this OPE to \emph{all-orders} in both the holomorphic and anti-holomorphic degrees of freedom: in momentum space, this means that we obtain all-order collinear expansions, while in CCFT this results in all-order expressions for the celestial OPE with all regular terms included. The resulting expressions are organized in terms of soft gluon descendants.


\subsection{Soft descendants}

In this section, we identify the soft gluon descendants used to organize the terms in the exact OPE between two vertex operators. The only non-trivial Wick contractions between gluon vertex operators (positive or negative helicity) are through the Kac-Moody current. While the singular part of the OPE is given by \eqref{CSOPE}, to obtain all-order expressions we require the \emph{exact} OPE between two Kac-Moody currents. This is obtained in the usual way, by expanding the current $j^{\msf{a}}(z)$ in modes (cf., \cite{Ketov:1995yd,DiFrancesco:1997nk,Blumenhagen:2009zz}): 
\begin{equation}\label{jj}
    j^{\msf{a}}(z_i)\,j^{\msf{b}}(z_j) = \frac{f^{\msf{abc}}\,j^{\msf{c}}(z_j)}{z_{ij}} + \sum_{n=0}^{\infty}\frac{z_{ij}^n}{n!}:\!\p^nj^{\msf{a}}\,j^{\msf{b}}\!:(z_j)\,,
\end{equation}
where $\partial$ denotes holomorphic differentiation and normal ordering $:\,:$ is defined by
\be\label{normalorder}
:\!\p^nj^{\msf{a}}\,j^{\msf{b}}\!:(z_j)=\frac{1}{2\pi\im}\oint_{C(z_j)}\d z_i\,\frac{\p^n j^{\msf{a}}(z_i)\, j^{\msf{b}}(z_j)}{z_{ij}}\,,
\ee
for $C(z_j)$ a small contour enclosing $z_j$. From now on, normal ordering with this prescription will be implicitly assumed.

To define soft gluon descendants of the vertex operators, we first identify the Kac-Moody descendants through the OPE between the Kac-Moody current and a gluon vertex operator (of either helicity) on the celestial sphere: 
\be\label{KMexpand}
j^{\msf{a}}(z_i)\,\cU^{\msf{b}}_{\pm,j}(z_j,\tilde{\kappa}_j) = \frac{f^{\msf{a}\msf{b}\msf{c}}\,\cU^{\msf{c}}(z_j,\tilde\kappa_j)}{z_{ij}} + \sum_{n=1}^\infty z_{ij}^{n-1}\,j^{\msf{a}}_{-n}\,\cU^{\msf{b}}_{\pm,j}(z_j,\tilde\kappa_j)\,,
\ee
where $\{j^{\msf{a}}_{-n}\}$ are the modes of the Kac-Moody current. The action of these modes on vertex operators is easily identified by comparing \eqref{KMexpand} and \eqref{jj}.
Now, conformally soft gluon currents on the celestial sphere are encoded in twistor space by polynomials in the two components of $\mu^{\dal}$, along with the Kac-Moody current~\cite{Adamo:2021lrv,Adamo:2021zpw}: 
\be\label{softcurrents}
J^{\msf{a}}[k,l](z) := \im^{k+l}\,j^{\msf{a}}(z)\,\big(\mu^{\dot1}\big)^k(z)\,\big(\mu^{\dot2}\big)^l(z)\,.
\ee
Note that $j^{\msf{a}} \equiv J^{\msf{a}}[0,0]$ in this decomposition. Soft gluon vertex operators are now defined by 
\be\label{softgluons}
J^{\msf{a}}[p](z,\tilde\kappa):= \frac{\im^{p}}{p!}\,j^{\msf{a}}(z)\,[\mu(z)\,\tilde\kappa]^p = \sum_{k+l=p}\frac{\tilde\kappa_{\dot1}{}^k\,\tilde\kappa_{\dot2}{}^l}{k!\,l!}\,J^{\msf{a}}[k,l]\,,\qquad p\in\Z_{\geq0}\,,
\ee
with the second equality following from the expansion of $[\mu\,\tilde\kappa]^p=(\mu^{\dot1}\tilde\kappa_{\dot1}+\mu^{\dot2}\tilde\kappa_{\dot2})^p$ using the binomial theorem. Acting with these soft gluons on a momentum eigenstate gives OPEs of the form
\begin{equation}\label{soft_descendants}
    J^{\msf{a}}[p](z_i,\tilde\kappa_i)\,\cU^{\msf{b}}_{\pm,j}(z_j,\tilde\kappa_j) = \frac{[i\,\p_j]^p}{p!}\,\frac{f^{{\msf{a}}{\msf{b}}\msf{c}}\,\cU^{{\msf{c}}}_{\pm,j}(z_j,\tilde\kappa_j)}{z_{ij}} +  \sum_{n=1}^\infty z_{ij}^{n-1}\,J_{-n}^{\msf{a}}[p](\tilde\kappa_i)\,\cU^{\msf{b}}_{\pm,j}(z_j,\tilde\kappa_j)\,,
\end{equation}
where $[i\,\p_j]\equiv\tilde\kappa_i^{\dal}\,\p/\p\tilde\kappa_j^{\dal}$. The coefficients $J_{-n}^{\msf{a}}[p](\tilde\kappa_i)$ are the \emph{soft gluon descendants} in terms of which all OPEs between the gluon vertex operators can be organized.

Using \eqref{softgluons} and \eqref{soft_descendants}, along with the basic OPEs of the Kac-Moody currents, it is straightforward to identify the action of the soft gluon descendants at each order. For instance, the action of the first soft descendant $J^{\msf{a}}_{-1}[1](\tilde{\kappa}_i)$ is given by 
\be\label{JV1}
J^{\msf{a}}_{-1}[1](\tilde{\kappa}_i)\,\cU^{\msf{b}}_{\pm,j}=\im\,f^{\msf{abc}}\,j^{\msf{c}}(z_j)\,[\partial\mu(z_j)\,i]\,\e^{\im\,[\mu(z_j)\,j]}+\im\,j^{\msf{a}}\,j^{\msf{b}}(z_j)\,[\mu(z_j)\,i]\,\e^{\im\,[\mu(z_j)\,j]}\,,
\ee
More generally, by writing any exact OPE as a series in $z_{ij}$ and comparing against the soft descendant expansion one can identify the action of the soft descendants at each order in the OPE expansion.

Now that we have the definition of soft descendants in hand, we are ready to write the OPEs between vertex operators in terms of them.


\subsection{The OPE in momentum space}\label{sec_momentum_space}

Armed with the soft descendant expansion, we can now consider the OPE between two positive helicity gluon vertex operators, in a momentum eigenstate basis, on the celestial sphere. The strategy is to perform the computation by writing all terms in the OPE as a series in $z_{ij}$, then recasting terms at each order in terms of combinations of soft descendants.

We start by computing the OPE of two positive helicity vertex operators \eqref{effective_vertex_operator_+}; using the Kac-Moody OPE \eqref{jj} and the fact that the $\mu^{\dot\alpha}(z)$ fields have no singular OPEs between themselves, it follows that
\begin{align}\label{VVope}
\cU_{+,i}^{\msf{a}}(z_i,\tilde\kappa_i)\,\cU_{+,j}^{\msf{b}}(z_j,\tilde\kappa_j) &= j^\msf{a}(z_i)\,j^\msf{b}(z_j)\,\e^{\im\,[\mu(z_i)\,i]}\,\e^{\im\,[\mu(z_j)\,j]}\nonumber \\
&=\left(\sum_{m=0}^{\infty} z_{ij}^{m-1}\,j^{\msf{a}}_{-m}\,j^{\msf{b}}\right)\left(\sum_{n=0}^{\infty}\frac{z_{ij}^n}{n!}\,\partial^n\!\left(\e^{\im\,[\mu\,i]}\right)\,\e^{\im\,[\mu\,j]}\right)\,,
\end{align}
where 
\be\label{expmodes}
j^{\msf{a}}_{-m}\,j^{\msf{b}}\vcentcolon=
\begin{cases}
f^{\msf{abc}}\,j^{\msf{c}}(z_j) & \mbox{when } m=0 \\
\vspace{-1em}\\
\displaystyle\frac{\partial^{m-1}j^{\msf{a}}\,j^{\msf{b}}(z_j)}{(m-1)!} & \mbox{when } m\geq 1
\end{cases}
\ee
$\partial\equiv\p/\p z_j$ and all fields on the second line of \eqref{VVope} are now evaluated at $z_j$ and implicitly normal-ordered. In writing this expression, we have only assumed that the fields $\mu^{\dal}(z)$ are analytic away from $z=0$ when Taylor expanding $\e^{\im\,[\mu(z_i)\,i]}$ around $z_j$. Schematically, \eqref{VVope} takes the form of an expansion in the separation $z_{ij}$:
\be
\cU_{+,i}^{\msf{a}}(z_i,\tilde\kappa_i)\,\cU_{+,j}^{\msf{b}}(z_j,\tilde\kappa_j)  = \sum_{p=0}^\infty z_{ij}^{p-1}\,U^{\msf{a}\msf{b}}_p(z_j,\tilde\kappa_i,\tilde\kappa_j)\,,
\ee
where the coefficients $U^{\msf{a}\msf{b}}_p$ are what we want to determine. For instance, the coefficient at $O(z_{ij}^{-1})$ (i.e., the leading, singular part of the OPE) is
\be\label{U-1}
U_{0}^{\msf{a}\msf{b}} = f^{\msf{a}\msf{b}\msf{c}}\,j^\msf{c}\,\e^{\im\,[\mu\,i]+\im\,[\mu\,j]} = f^{\msf{abc}}\,\cU_+^\msf{c}(z_j,\tilde\kappa_i+\tilde\kappa_j)\,,
\ee
which corresponds to the well-known collinear splitting function for positive helicity gluons~\cite{Altarelli:1977zs,Birthwright:2005ak}.

\medskip

From \eqref{VVope}, one identifies
\be
U_p^\msf{ab}(z_j,\tilde{\kappa}_i,\tilde{\kappa}_j) = \sum_{n=0}^p\frac{1}{n!}\,j^\msf{a}_{n-p}\,j^b\,\e^{\im\,[\mu\,j]}\,\p^n\!\left(\e^{\im\,[\mu\,i]}\right)\,.
\ee
In order to evaluate $\p^n(\e^{\im\,[\mu\,i]})$, one needs the following main identity: for any smooth $f(z)$, the $n^\text{th}$ derivative $\p^n\e^{f(z)}$ of its exponential is given by:
\be\label{dnef0}
\p^n\e^{f(z)} = e^{f(z)}\,\sum_{\ell=0}^n\sum_{k=0}^\ell\frac{(-1)^{\ell-k}}{k!\,(\ell-k)!}\,f^{\ell-k}(z)\,\p^nf^{k}(z)\,.
\ee
We review a brief proof of this identity in appendix~\ref{app:exp}. Applying it to $f(z)=\im\,[\mu(z)\,i]$ gives
\be\label{Upab}
U_p^\msf{ab} = \sum_{n=0}^p\frac{1}{n!}\,j^\msf{a}_{n-p}\,j^{\msf{b}}\,\e^{\im\,[\mu\,i]+\im\,[\mu\,j]}\,\sum_{\ell=0}^n\sum_{k=0}^\ell\frac{(-1)^{\ell-k}\,\im^\ell}{k!\,(\ell-k)!}\,[\mu\,i]^{\ell-k}\,\p^n\left([\mu\,i]^{k}\right)\,.
\ee
A second useful identity, valid for any smooth $f(z)$, states that:
\be
\sum_{k=0}^\ell\frac{(-1)^{\ell-k}}{k!(\ell-k)!}\,f^{\ell-k}\,\p^nf^k = 0\,,\qquad \forall\,\ell>n\,,
\ee
see equation \eqref{goodboi} of appendix \ref{app:exp}. Using this for $f(z)=\im\,[\mu(z)\,i]$, the second sum in \eqref{Upab} can be trivially extended from $0\leq\ell\leq n$ to $0\leq\ell\leq p$, since the summands for $n+1\leq\ell\leq p$ vanish identically:
\be\label{Upab1}
U_p^\msf{ab} = \sum_{n=0}^p\frac{1}{n!}\,j^\msf{a}_{n-p}\,j^{\msf{b}}\,\e^{\im\,[\mu\,i]+\im\,[\mu\,j]}\sum_{\ell=0}^p\sum_{k=0}^\ell\frac{(-1)^{\ell-k}\,\im^\ell}{k!\,(\ell-k)!}\,[\mu\,i]^{\ell-k}\,\p^n\!\left([\mu\,i]^{k}\right)\,.
\ee
As a result, the sum over $n$ can be commuted through the other sums. This expression is now ready to be recast in terms of soft descendants of the gluon operator $\cU_+^\msf{c}(z,\tilde\kappa_1+\tilde\kappa_2)$.

\medskip

The OPE of a hard gluon $\cU^\msf{b}_{+,j}$ with a soft gluon $J^\msf{a}[k](z_i,\tilde\kappa_i) = j^\msf{a}(z_i)\,[\mu(z_i)\,i]^k$ can be written out along the sames lines as \eqref{VVope}, giving:
\begin{align}
J^\msf{a}[k](z_i,\tilde\kappa_i)\,\cU_{+,j}^\msf{b}(z_j,\tilde\kappa_j) &= \frac{\im^k}{k!}\,j^\msf{a}(z_i)\, j^\msf{b}(z_j)\,[\mu(z_i)\,i]^k\,\e^{\im\,[\mu(z_j)\,j]}\nonumber\\
&= \frac{\im^k}{k!}\left(\sum_{m=0}^{\infty}z_{ij}^{m-1}\,j^\msf{a}_{-m}\,j^\msf{b}\right)\left(\sum_{n=0}^{\infty}\frac{z_{ij}^n}{n!}\,\e^{\im\,[\mu\,j]}\,\p^n\!\left([\mu\,i]^k\right)\right)\,.
\end{align}
Comparing with \eqref{soft_descendants}, we see that a typical soft descendant has the explicit expression
\be
J^\msf{a}_{-p}[k](\tilde\kappa_i)\,\cU_{+,j}^\msf{b}(z_j,\tilde\kappa_j) = \frac{\im^k}{k!}\,\sum_{n=0}^p\frac{1}{n!}\,j^\msf{a}_{n-p}\,j^\msf{b}\,\p^n\!\left([\mu\,i]^k\right)\,\e^{\im\,[\mu\,j]}\,.
\ee
By shifting $\tilde\kappa_j\mapsto\tilde\kappa_j+\tilde\kappa_i$ in this expression, one also obtains
\be\label{softdesc}
J^\msf{a}_{-p}[k](\tilde\kappa_i)\,\cU_+^\msf{b}(z_j,\tilde\kappa_i+\tilde\kappa_j) = \frac{\im^k}{k!}\,\sum_{n=0}^p\frac{1}{n!}\,j^\msf{a}_{n-p}\,j^\msf{b}\,\p^n\!\left([\mu\,i]^k\right)\,\e^{\im\,[\mu\,i]+\im\,[\mu\,j]}\,,
\ee
which contains precisely the kind of terms entering \eqref{Upab1}.

Performing the sum over $n$ in \eqref{Upab1} using \eqref{softdesc} gives all of the expansion coefficients in the OPE between two positive helicity gluons in a basis of soft gluon descendants:
\be\label{Upab2}
U_p^\msf{ab} = \sum_{\ell=0}^p\sum_{k=0}^\ell\frac{(-\im\,[\mu\,i])^{\ell-k}}{(\ell-k)!}\,J^\msf{a}_{-p}[k](\tilde\kappa_i)\,\cU_+^\msf{b}(z_j,\tilde\kappa_i+\tilde\kappa_j)\,.
\ee
Written in terms of these coefficients, we now obtain an all-order expression for the celestial OPE
\be
\cU^\msf{a}_{+,i}\,\cU^\msf{b}_{+,j} = \sum_{p=0}^{\infty}z_{ij}^{p-1}\sum_{\ell=0}^p\sum_{k=0}^\ell\frac{(-[i\,\p_j])^{\ell-k}}{(\ell-k)!}\,J^\msf{a}_{-p}[k](\tilde\kappa_i)\,\cU_+^\msf{b}(z_j,\tilde\kappa_i+\tilde\kappa_j)\,,
\ee
where factors of $\im\,[\mu\,i]$ have been replaced with the differential operators
\be\label{opreplace}
\im\,[\mu\,i]\longleftrightarrow[i\,\p_j]\equiv\tilde\kappa_i^{\dal}\frac{\p}{\p\tilde\kappa_j^{\dal}}\,,
\ee
when acting on $\e^{\im\,[\mu\,i]+\im\,[\mu\,j]}$. 

Exchanging the sums over $\ell$ and $k$, then shifting $\ell\mapsto\ell+k$ simplifies the result further to
\be\label{++mom}
\boxed{\cU^\msf{a}_{+,i}\,\cU^\msf{b}_{+,j} = \sum_{p=0}^{\infty}z_{ij}^{p-1}\sum_{k=0}^p\sum_{\ell=0}^{p-k}\frac{(-[i\,\p_j])^{\ell}}{\ell!}\,J^\msf{a}_{-p}[k](\tilde\kappa_i)\,\cU_+^\msf{b}(z_j,\tilde\kappa_i+\tilde\kappa_j)\,.}
\ee
This is the exact tree-level celestial OPE, to all orders in $z_{ij}$, for two positive helicity gluons in the MHV sector in a momentum eigenstate basis. Note that this expression also encodes all orders in the anti-holomorphic collinear limit as well: for each $p\geq0$, the soft descendant $J^\msf{a}_{-p}[k](\tilde\kappa_i)\,\cU_+^\msf{b}(z_j,\tilde\kappa_i+\tilde\kappa_j)$ can be Taylor expanded around $\tilde{\kappa}_i=\tilde{\kappa}_j$. Since this anti-holomorphic dependence is exponential, this expansion generates all-order terms in $[i\,j]$ as well.

\medskip

The calculation of the celestial OPE proceeds along identical lines for the mixed-helicity configuration, involving one positive and one negative helicity gluon. In this case, the only difference is the insertion of $\msf{O}(z)$ -- built from the twistor fermions -- which accompanies the negative helicity vertex operator \eqref{effective_vertex_operator_-}. The end result is a celestial OPE
\be\label{+-mom}
\boxed{\cU^\msf{a}_{+,i}\,\cU^\msf{b}_{-,j} = \sum_{p=0}^{\infty}z_{ij}^{p-1}\sum_{k=0}^p\sum_{\ell=0}^{p-k}\frac{(-[i\,\p_j])^{\ell}}{\ell!}\,J^\msf{a}_{-p}[k](\tilde\kappa_i)\,\cU_-^\msf{b}(z_j,\tilde\kappa_i+\tilde\kappa_j)\,,}
\ee
with all the resulting terms proportional to soft gluon descendants of a \emph{negative} helicity gluon.

At first, it may seem that the result is incomplete: the leading collinear splitting function for mixed-helicity gluons includes two terms, one for each helicity, whereas the $p=0$ term in \eqref{+-mom} contains only the negative helicity contribution to this splitting function. The key is to remember that \eqref{+-mom} holds only for the MHV sector, where positive helicity terms in the mixed-helicity OPE are identically zero. Phrased differently, the MHV amplitude is homogeneous with respect to collinear limits (i.e., all collinear limits of a MHV amplitude give lower-point MHV amplitudes), and positive helicity terms on the right-hand-side of \eqref{+-mom} would violate this homogeneity. Indeed, such terms would correspond to generating a tree-level gluon scattering amplitude with only one negative helicity particle, which are identically zero.

\medskip

Finally, one can consider the OPE between two negative helicity vertex operators. This OPE is on a different footing to the others due to the inherent chirality of the MHV sector; this is captured in the fact that the OPE between $\cU^{\msf{a}}_{-,i}$ and $\cU^{\msf{b}}_{-,j}$ is proportional to
\be\label{--ope1}
\msf{O}(z_i)\,\msf{O}(z_j)=\msf{O}(z_j)\,\sum_{m=0}^{\infty}\frac{z_{ij}^m}{m!}\,\partial^{m}\msf{O}(z_j)\,,
\ee
where $\msf{O}(z)\equiv \chi^{4}(z)$. Since 
\be
\chi^{a}(z)=\theta^{a0}+z\,\theta^{a1}\,,
\ee
and there are only 8 degrees of freedom in the fermionic $\theta^{a\alpha}$, only the $m=4$ term from the expansion \eqref{--ope1} survives, consistent with \eqref{Grassint}. The rest of the OPE calculation proceeds as before, leaving:
\be\label{--ope2}
\cU^{\msf{a}}_{-,i}\,\cU^{\msf{b}}_{-,j}=\frac{\theta^{4}}{4!}\,\sum_{p=0}^{\infty}z_{ij}^{p+3}\sum_{k=0}^p\sum_{\ell=0}^{p-k}\frac{(-[i\,\p_j])^{\ell}}{\ell!}\,J^\msf{a}_{-p}[k](\tilde\kappa_i)\,\cU_-^\msf{b}(z_j,\tilde\kappa_i+\tilde\kappa_j)\,,
\ee
where 
\be\label{thetaab}
\theta^{4}:=\frac{\epsilon_{abcd}}{4!}\theta^{a1}\,\theta^{b1}\,\theta^{c1}\,\theta^{d1}\,,
\ee
denotes the combination of fermionic moduli produced by the expansion.

Note that this OPE has no singular term, as expected for the MHV sector. If this OPE were singular, it would correspond to generating a collinear limit of the tree-level MHV amplitude which had one negative helicity gluon, and such tree amplitudes are identically zero. This is tied to the presence of $\theta^4$ in \eqref{--ope2}, which indicates that the OPE only makes sense in the context of the complete correlator \eqref{EFTcorr}, where the $\d^8\theta$ integral must be saturated. Regularity of this expression is also consistent with generating non-trivial form factors for two anti-self-dual fields in self-dual Yang-Mills theory~\cite{Costello:2022wso}.

Hence, the expression \eqref{--ope2} is not a self-contained celestial OPE, since it cannot be written purely in terms of locations on the celestial sphere and descendants of the vertex operators. This is due to the underlying chirality of the twistor string, for which the OPE between negative helicity vertex operators is not really well defined. As such, \eqref{--ope2} will not provide good OPE data in the context of CCFT, and we will not consider this helicity configuration in the MHV sector any further.  


\subsection{The OPE in Celestial CFT}

By Mellin transforming the results \eqref{++mom} and \eqref{+-mom} to the conformal primary basis, we can now obtain all-order celestial OPEs in the context of CCFT, which include all regular contributions in the MHV sector. To do this, we follow the rule \eqref{cVOs} to implement the Mellin transform while making all dependence on the frequencies explicit using \eqref{momparam} and \eqref{softdesc}. In doing this, it is easier to replace $[i\,\partial_j]$ with $\im\,[\mu\,i]$ according to \eqref{opreplace} at the beginning of the calculation. 

Starting with the OPE between positive helicity gluons, the Mellin transformed OPE is
\begin{multline}\label{MT1}
\cU_{+,\Delta_i}^{\msf{a},\veps_i}(z_i,\bar z_i)\,\cU_{+,\Delta_j}^{\msf{b},\veps_j}(z_j,\bar z_j)=\int_{\R_+^2}\d\omega_i\,\omega_i^{\Delta_i-2}\,\d\omega_j\,\omega_j^{\Delta_j-2}\sum_{p=0}^{\infty}z_{ij}^{p-1} \\
\times\sum_{k=0}^{p}\sum_{\ell=0}^{p-k}(\varepsilon_i\,\omega_i)^{\ell}\,\frac{(-\im\,[\mu\,\bar{z}_i])^\ell}{\ell!}\,(\varepsilon_i\,\omega_i)^{k}
\, J^{\msf{a}}_{-p}[k](\bar{z}_i)\,j^{\msf{b}}\,\e^{\im\veps_i\omega_i[\mu\bar z_i]+\im\veps_j\omega_j[\mu\bar z_j]}\,,
\end{multline}
where all fields are evaluated at $z_j$ with implicit normal-ordering on the right-hand-side of the OPE relation, and we have abbreviated $\bar z_{ij\,\dal}\equiv\bar z_{i\,\dal}-\bar z_{j\,\dal} = (0,\bar z_{ij})$ (with $\bar z_{ij}=\bar z_i-\bar z_j$ as usual). Now, the argument of the exponential in this expression can be conveniently rewritten as
\begin{equation}
    \im\,\veps_i\,\omega_i\,[\mu\,\bar z_i]+\im\,\veps_j\,\omega_j\,[\mu\,\bar z_j] = \im\,(\veps_i\omega_i+\veps_j\omega_j)\left([\mu\,\bar z_j] + \frac{[\mu\,\bar z_{ij}]}{1+\frac{\veps_j\omega_j}{\veps_i\omega_i}}\right)\,,
\end{equation}
and combined with rescaling the first Mellin integral by $\omega_i\rightarrow\omega_i\omega_j$ this leaves
\begin{multline}
\cU_{+,\Delta_i}^{\msf{a},\veps_i}\,\cU_{+,\Delta_j}^{\msf{b},\veps_j} = \sum_{p=0}^{\infty}z_{ij}^{p-1}\sum_{k=0}^p\sum_{\ell=0}^{p-k} \int_{\mathbb{R}_+^2} \d\omega_i\,\omega_i^{\ell+k+\Delta_i-2}\, \d\omega_j\,\omega_j^{k+\ell+\Delta_i+\Delta_j-3}\\
   \times\, \veps_i^{\ell+k}\,\frac{(-\im\,[\mu\,\bar{z}_i])^\ell}{\ell!}\;J_{-p}^\msf{a}[k](\bar z_i)\, j^{\msf{b}}\,\text{exp}\bigg[{\im\,(\veps_i\omega_i+\veps_j)\,\omega_j\left([\mu\,\bar z_j] + \frac{[\mu\,\bar z_{ij}]}{1+\frac{\veps_j}{\veps_i\omega_i}}\right)}\bigg]\,.
\end{multline}
After rescaling the second Mellin integral by $\omega_j \rightarrow \frac{\omega_j}{|\veps_j+\veps_i\omega_i|}$, it is possible to Taylor expand the exponential in $\bar{z}_{ij}$:
\begin{multline}\label{inter1}
\cU_{+,\Delta_i}^{\msf{a},\veps_i}\,\cU_{+,\Delta_j}^{\msf{b},\veps_j}= \sum_{p=0}^{\infty}z_{ij}^{p-1}\sum_{k=0}^p\sum_{\ell=0}^{p-k}\sum_{m=0}^{\infty} \int_{\mathbb{R}_+^2} \frac{\d\omega_i\,\omega_i^{\ell+k+m+\Delta_i-2}}{|\veps_j+\veps_i\omega_i|^{\Delta_i+\Delta_j+k+\ell+m-2}}\,\d\omega_j\,\omega_j^{\Delta_i+\Delta_j+k+\ell-3} \\
    \times\,\veps_i^{\ell+k+m}\,[\text{sgn}(\veps_i\omega_i+\veps_j)]^{m}\,J_{-p}^\msf{a}[k](\bar z_i)\,j^{\msf{b}}\,\frac{\bar z_{ij}^{m}}{m!}\,\frac{(-\im\,[\mu\,\bar{z}_i])^{\ell}}{\ell!} \\
    \times\,\bar\partial^m_j \,\text{exp}\bigg[{\im\,\text{sgn}(\veps_i\omega_i+\veps_j)\,\omega_j\,[\mu\,\bar{z}_j] }\bigg]\,,
\end{multline}
where $\dbar_j\equiv\frac{\p}{\p\bar{z}_j}$. Already, we see that the expression explicitly contains all orders in both $z_{ij}$ \emph{and} $\bar{z}_{ij}$.

Using the fact that $[\mu\,\bar{z}_i]=\mu^{\dot{0}}+\bar{z}_i\,\mu^{\dot{1}}$, a short calculation demonstrates that
\be\label{opreplace2}
-\im\,[\mu\,\bar{z}_i]\longleftrightarrow \frac{1}{\omega_j\,\text{sgn}(\veps_i\,\omega_i+\veps_j)}\left(-\bar{z}_{ij}\,\dbar_j-\omega_j\,\frac{\partial}{\partial\omega_j}\right)\,,
\ee
when acting on the exponential in \eqref{inter1}. The Euler vector $\omega_j\,\frac{\partial}{\partial\omega_j}$ can then be integrated-by-parts to extract the homogeneity in $\omega_j$ of all non-exponential factors in each term of the expression. This results in:
\begin{multline}
\cU_{+,\Delta_i}^{\msf{a},\veps_i}\,\cU_{+,\Delta_j}^{\msf{b},\veps_j}= \sum_{p=0}^{\infty}z_{ij}^{p-1}\sum_{k=0}^{p}\sum_{\ell=0}^{p-k}\sum_{m=0}^{\infty}\frac{\bar{z}^m_{ij}}{m!}\,\veps_i^{m+\ell+k}\,\frac{\bar{D}_k^{\ell}}{\ell!}\,J^{\msf{a}}_{-p}[k](\bar{z}_i)\,j^{\msf{b}}\,\bar\partial^m_j \\
\times \int_{\mathbb{R}_+^2}\frac{\d\omega_i\,\omega_i^{\ell+k+m+\Delta_i-2}\,\d\omega_j\,\omega_j^{\Delta_i+\Delta_j+k-3}}{\left|1+\frac{\veps_i\omega_i}{\veps_j}\right|^{\Delta_i+\Delta_j+k+\ell+m-2}}\,[\text{sgn}(\veps_i\,\omega_i+\veps_j)]^{m+\ell}\,\e^{\im\,\text{sgn}(\veps_i\omega_i+\veps_j)\,\omega_j\,[\mu\,\bar{z}_j] }\,,
\end{multline}
where
\be
\bar{D}_{k}:=-\bar{z}_{ij}\,\dbar_j+\Delta_i+\Delta_j+k-3\,,
\ee
is a differential operator on the celestial sphere, dependent on the conformal dimensions as well as the summation index $k$.

At this stage, we observe that 
\begin{equation}
\int_{0}^{\infty}\!\d\omega_j\,\omega_j^{\Delta_i+\Delta_j+k-3}\,j^{b}(z_j)\,\text{exp}\bigg[{\im\,\text{sgn}(\veps_i\omega_i+\veps_j)\,\omega_j\,[\mu\,\bar{z}_j] }\bigg]=\cU^{\msf{b},\mathrm{sgn}(\veps_i\omega_i+\veps_j)}_{+,\Delta_i+\Delta_j+k-1}(z_j,\bar{z}_j)\,,
\end{equation}
by virtue of \eqref{cVOs}. Combined with some trivial algebraic rearrangements, we are left with the all-orders celestial OPE in the conformal primary basis
\begin{equation}\label{++_final}
\boxed{
\begin{aligned}
 \cU&_{+,\Delta_i}^{\msf{a},\veps_i}(z_i,\bar z_i)\,\cU_{+,\Delta_j}^{\msf{b},\veps_j}(z_j,\bar z_j)= \sum_{p=0}^{\infty}z_{ij}^{p-1}\sum_{k=0}^p\sum_{\ell=0}^{p-k}\sum_{m=0}^{\infty}\frac{\bar z_{ij}^m}{m!} \,\veps_i^{m+\ell+k}\,\frac{\bar{D}_k^{\ell}}{\ell !}\, J^{\msf{a}}_{-p}[k](\bar z_i)\\
    &\times\,\int\limits_{0}^{\infty} \frac{\d\omega_i\,\omega_i^{\ell+k+m+\Delta_i-2}}{\left|\frac{\veps_i\omega_i}{\veps_j}+1\right|^{\Delta_i+\Delta_j+k+\ell+m-2}}\,[\text{sgn}(\veps_i\,\omega_i+\veps_j)]^{m+\ell}
    \,\bar\partial^m_j\,\cU_{+,\Delta_i+\Delta_j+k-1}^{\msf{b}, \text{sgn}(\veps_i\omega+\veps_j)}(z_j,\bar z_j)\,.
\end{aligned}}
\end{equation}
This provides a `master' formula for the celestial OPE, valid for any configuration of incoming/outgoing positive helicity gluons in the MHV sector. 

It is easy to see that the $p=0$ terms in \eqref{++_final} correctly reproduce the master formula for the singular contribution to the OPE~\cite{Adamo:2021zpw}. It is illustrative to evaluate the OPE for explicit incoming/outgoing configurations, where the remaining Mellin frequency integral can be performed explicitly. For instance, when both gluons are in the same configuration ($\veps_i=\veps_j=\veps$), one obtains
\begin{multline}
\cU_{+,\Delta_i}^{\msf{a},\veps}\,\cU_{+,\Delta_j}^{\msf{b},\veps}= \sum_{p=0}^{\infty}z_{ij}^{p-1}\sum_{k=0}^p\sum_{\ell=0}^{p-k}\sum_{m=0}^{\infty}\frac{\bar z_{ij}^m}{m!}\,\frac{\bar{D}_k^{\ell}}{\ell !}\, J^{\msf{a}}_{-p}[k](\bar z_i)\\
    \times\,B(\Delta_i+k+\ell+m-1,\,\Delta_j-1)\,\bar\partial^m_j\,\cU_{+,\Delta_i+\Delta_j+k-1}^{\msf{b},\veps}(z_j,\bar z_j)\,,  
\end{multline}
where $B(x,y)$ denotes the Euler Beta function. The $p=0$ terms in this expression contain the single tower of SL$(2,\R)$ descendants that come with the pole in $z_{ij}$ which were previously determined in the literature~\cite{Fan:2019emx,Pate:2019lpp,Adamo:2021zpw}; while these singular terms are valid in general, all terms for $p>0$ hold only for the celestial OPE within the MHV sector of gluon scattering.

\medskip

The Mellin transform of the celestial OPE between a positive and a negative helicity gluon \eqref{+-mom} follows similar lines; the only difference is in the initial definition of the transform with respect to the frequency of the negative helicity particle via \eqref{cVOs}. The resulting master formula is:
\begin{equation}\label{+-_final}
\boxed{
\begin{aligned}
\cU&_{+,\Delta_i}^{\msf{a},\veps_i}(z_i,\bar z_i)\,\cU_{-,\Delta_j}^{\msf{b},\veps_j}(z_j,\bar z_j)= \sum_{p=0}^{\infty}z_{ij}^{p-1}\sum_{k=0}^p\sum_{\ell=0}^{p-k}\sum_{m=0}^{\infty}\frac{\bar z_{ij}^m}{m!} \,\veps_i^{m+\ell+k}\,\frac{\bar{D}_k^{\prime\,\ell}}{\ell !}\, J^{\msf{a}}_{-p}[k](\bar z_i)\\
    &\times\,\int\limits_{0}^{\infty} \frac{\d\omega_i\,\omega_i^{\ell+k+m+\Delta_i-2}}{\left|\frac{\veps_i\omega_i}{\veps_j}+1\right|^{\Delta_i+\Delta_j+k+\ell+m}}\,[\text{sgn}(\veps_i\,\omega_i+\veps_j)]^{m+\ell}
    \,\bar\partial^m_j\,\cU_{-,\Delta_i+\Delta_j+k-1}^{\msf{b}, \text{sgn}(\veps_i\omega+\veps_j)}(z_j,\bar z_j)\,,
\end{aligned}}
\end{equation}
where
\be\label{primeop}
\bar{D}_k^{\prime}:=-\bar{z}_{ij}\,\dbar_j+\Delta_i+\Delta_j+k-1\,,.
\ee
As in momentum space, this OPE features only contributions from soft gluon descendants of the negative helicity gluon on the celestial sphere. Indeed, the $p=0$ terms in \eqref{+-_final} only contain one (chiral) half of those appearing in the master formula for the singular, mixed helicity celestial OPE~\cite{Adamo:2021zpw}. This is, once again, an intrinsic feature of the mixed helicity OPE in the MHV sector (as discussed above). Whereas the singular terms in the same helicity OPE \eqref{++_final} are valid beyond the MHV sector, even the singular terms in the mixed helicity OPE are sensitive to the helicity configuration on the celestial sphere.

For the sake of concreteness, one can consider the mixed helicity OPE for gluons in the same incoming/outgoing configuration ($\veps_i=\veps_j=\veps$). This leads to an explicit all-orders tower of OPE coefficients
\begin{multline}
\cU_{+,\Delta_i}^{\msf{a},\veps}\,\cU_{-,\Delta_j}^{\msf{b},\veps}= \sum_{p=0}^{\infty}z_{ij}^{p-1}\sum_{k=0}^p\sum_{\ell=0}^{p-k}\sum_{m=0}^{\infty}\frac{\bar z_{ij}^m}{m!}\,\frac{\bar{D}_k^{\prime\,\ell}}{\ell !}\, J^{\msf{a}}_{-p}[k](\bar z_i)\\
    \times\,B(\Delta_i+k+\ell+m-1,\,\Delta_j+1)\,\bar\partial^m_j\,\cU_{-,\Delta_i+\Delta_j+k-1}^{\msf{b},\veps}(z_j,\bar z_j)\,,  
\end{multline}
defined in terms of the Euler Beta function. It is easy to see that the $p=0$ terms in this expression coincide with one chiral half of the singular terms for one SL$(2,\R)$ descendant tower in the OPE of mixed helicity celestial gluons which have been appeared previously~\cite{Pate:2019lpp,Adamo:2021zpw}. 


\subsection{Null states}

Comparing our results with previous expressions at subleading level in the literature~\cite{Ebert:2020nqf} is an involved process. This is due to the fact that we use soft-descendants to organize the OPE expansion, while the expression in~\cite{Ebert:2020nqf} is written in terms of Kac-Moody and Virasoro descendants. The way to bridge this gap is to consider the existence of null states in CCFT. On these states, we expect that the action of the soft currents can be recast in terms of the Kac-Moody and Virasoro currents. That is, soft descendants of these states can be rewritten in a basis of Kac-Moody and Virasoro descendants.  

While these null relations are not \emph{a priori} known, we can leverage our free field realization of the target space algebra at MHV level to calculate them explicitly. This is rather trivial at first order. In this case the leading soft current is given by the residue around $\Delta=1$ of the hard particle $\mathcal{U}_{\Delta}$ (cf., \cite{Donnay:2018neh,Adamo:2019ipt,Adamo:2021zpw}). In our model, this is the residue of the vertex operator \eqref{cVOs} which is the Kac-Moody current $j^\msf{a}$ itself; this is just a restatement of \eqref{softcurrents}, where $J^\msf{a}[0,0]=j^\msf{a}$. The next soft current is more interesting. From \eqref{softcurrents}, it is a combination of the Kac-Moody current and the $\mu$ operators. In a conformal basis it can be written as
\begin{equation}
J^\msf{a}[1]=J^\msf{a}[1,0]+\bar{z}\,J^\msf{a}[0,1]=j^\msf{a}\,[\mu(z)\,\bar{z}]\,,
\end{equation}
and its descendant as 
\begin{equation}\label{nulldesc}
    J^\msf{a}_{-1}[1]=j^\msf{a}_0\,[\mu_{-1}\,\bar z]+j^\msf{a}_{-1}\,[\mu_{0}\,\bar z].
\end{equation}
The descendants of $\mu^{\dot{\alpha}}$ acting on a gluon operator are given by a Taylor expansion, since there is no short distance singularity:
\begin{equation}
    \mu^{\dot{\alpha}}(z_i)\,\cU^\msf{a}_\Delta(z_j)=\sum_{n=0}^{\infty}\frac{z^n_{ij}}{n!}\partial^n\mu^{\dot{\alpha}}\,\cU^\msf{a}_\Delta(z_j)\equiv\sum_{p=1}^{\infty}z^{p-1}_{ij}\,\mu_{-p}^{\dot{\alpha}}\,\cU^\msf{a}_\Delta(z_j)\,, 
\end{equation}
suppressing irrelevant helicity and incoming/outgoing labels. This means that 
\begin{equation}
[\mu_0\,\bar z]\,\cU^\msf{a}_\Delta(z)=[\mu\,\bar z]\,\cU^\msf{a}_\Delta(z)=(\Delta-2)\,\cU^\msf{a}_{\Delta-1}(z)\,,
\end{equation}
and
\begin{equation}
    [\mu_{-1}\,\bar z]\,\cU^\msf{a}_\Delta(z)=[\partial\mu\,\bar z]\,\cU^\msf{a}_\Delta(z)\,,
\end{equation}
in terms of the modes appearing in the soft gluon descendant \eqref{nulldesc}.

Using these relations, the action of $J^\msf{a}_{-1}[1]$ on a hard gluon in the conformal primary basis can be written as
\begin{equation}\label{null_state}
    J^\msf{a}_{-1}[1]\,\cU^{\msf{b}}_{\Delta}=(\Delta-1)\,j^\msf{a}_{-1}\,\cU^b_{\Delta-1}-j_{-1}^\msf{b}\,\cU^\msf{a}_{\Delta-1}-f^{\msf{a}\msf{b}\msf{c}}\,L_{-1}\,\cU^\msf{c}_{\Delta-1}\,,
\end{equation}
to lowest order in $\bar{z}$, where $L_{-1}=\partial$ is the SL$(2,\C)$/Virasoro generator. Equation \eqref{null_state} is the same as the null state condition of~\cite{Banerjee:2020zlg}, and can be used to rewrite the OPE at subleading order in a basis of Kac-Moody and Virasoro descendants matching the subleading terms found in~\cite{Ebert:2020nqf}. We expect that similar null relations hold for other soft current descendants. These can, in principle, be derived from the representation \eqref{softcurrents} for the soft currents and their action on hard particles using the effective OPEs, though the process quickly becomes rather cumbersome.


\section{Discussion}\label{Sec:Conc}

In this paper, we used twistor string theory to provide all-order expressions for tree-level celestial OPEs of gluons in the MHV sector. The results \eqref{++mom}, \eqref{+-mom}, \eqref{++_final} and \eqref{+-_final} are organized in terms of soft gluon descendants, when enables the OPE to close on the gluon vertex operators. These provide the first instances where all regular contributions to the celestial OPE, needed to constrain any putative CCFT, are captured in their entirety.

There are many interesting directions which can be explored following on from this work. While the twistor string provides a dynamical principle for generating the celestial OPEs, it is natural to ask if there are alternative ways to obtain our results using more standard amplitudes technology; we will return to this in future work \cite{renta}. It should also be possible to obtain all-order celestial OPEs for the MHV sector of \emph{graviton} scattering; we attempted to do this using the twistor string of~\cite{Skinner:2013xp} for $\cN=8$ supergravity, but ran into technical obstructions related to the worldsheet supersymmetry of that model. A more straightforward route is probably to use chiral twistor sigma models~\cite{Adamo:2021bej,Adamo:2021lrv,Sharma:2022arl}, which describe the MHV sector in terms of a classical 2d CFT on the celestial sphere directly.

In addition, one can ask to what extent it is possible to push these all-order results beyond the MHV sector. From the perspective of twistor string theory, it is unclear how to achieve this: N$^{k}$MHV scattering corresponds to degree $k+1$ holomorphic maps from the Riemann sphere to twistor space, so the worldsheet is no longer identified with the celestial sphere at generic points in the moduli space. While this identification does emerge in the strict OPE limit on the worldsheet~\cite{Adamo:2021zpw}, it only captures the singular contributions to the celestial OPE.

\acknowledgments

We thank Lecheng Ren, Anders Schreiber and Diandian Wang for useful discussions. TA is supported by a Royal Society University Research Fellowship and by the Leverhulme Trust (RPG-2020-386). WB is supported by a Royal Society PhD studentship. EC thanks the support of the Frankel-Goldfield-Valani Research Fund. AS is supported by a Black Hole Initiative fellowship, which is funded by the Gordon and Betty Moore Foundation and the John Templeton Foundation.

\appendix

\section{Useful identities}
\label{app:exp}

Let $\p\equiv\p/\p z$. In deriving the celestial OPE, we needed the functional identity
\be\label{dnef}
\p^n\e^f = \e^f\,\sum_{k=0}^n\sum_{j=0}^k\frac{(-f)^{k-j}\,\p^nf^{j}}{j!\,(k-j)!}\,,
\ee
that computes $z$ derivatives of $\e^{f(z)}$ for a given smooth function $f(z)$. This is a special case of Hoppe's formula for the derivatives of composite functions. In this appendix, we review a proof of this identity that is adapted from the more general presentation in~\cite{Johnson:2002} and references therein.

Consider the slightly more general problem of computing derivatives of $\e^{tf(z)}$ in the presence of an auxiliary parameter $t$. It is straightforward to inductively verify that the $n^\text{th}$ derivative $\p^n\e^{tf}$ is $\e^{tf}$ times a polynomial of degree $n$ in $t$,
\be\label{ansatz}
\p^n\e^{tf} = \e^{tf}\,\sum_{k=0}^n a_{n,k}\,t^k
\ee
where the coefficients $a_{n,k}$ depend on $f$ and its derivatives but not on $t$. If we can determine the $a_{n,k}$, we can compute derivatives of $\e^f$ by setting $t=1$.

In order to find $a_{n,k}$, we will relate these to the derivatives $\p^nf^j$ for $0\leq j\leq n$,
\be\label{pnfj}
    \p^nf^j = \p_t^j\bigl(\p^n\e^{tf}\bigr)\bigr|_{t=0} = \sum_{r=0}^na_{n,r}\,\p_t^j(t^r\e^{tf})\bigr|_{t=0}= \sum_{r=0}^j\frac{j!}{(j-r)!}\,f^{j-r}\,a_{n,r}
\ee
where we have substituted \eqref{ansatz} to evaluate the $t$ derivatives. These provide a system of algebraic equations for $a_{n,r}$. To invert them, we compute for $0\leq k\leq n$:
\begin{align}\label{finiteinv}
    \sum_{j=0}^k\frac{(-f)^{k-j}\p^nf^{j}}{j!(k-j)!} &= \sum_{j=0}^k\sum_{r=0}^j\frac{(-1)^{k-j}f^{k-r}a_{n,r}}{(k-j)!(j-r)!}\nonumber\\
    &= \sum_{r=0}^k\frac{(-f)^{k-r}a_{n,r}}{(k-r)!}\sum_{j=r}^k\binom{k-r}{j-r}\,(-1)^{j-r}\nonumber\\
    &= \sum_{r=0}^k\frac{(-f)^{k-r}a_{n,r}}{(k-r)!}\;(1-1)^{k-r}\nonumber\\
    &= a_{n,k}\,,
\end{align}
having exchanged the sums over $j$ and $r$ in the second line, and applied the binomial theorem in the third. Inserting this solution for $a_{n,k}$ in \eqref{ansatz} and setting $t=1$ gives us the required identity \eqref{dnef}.

A further useful identity needed for the simplification of the OPE is obtained by repeating this computation for $k>n$. The calculation of \eqref{pnfj} can be generalized to arbitrary $j\in\Z_{>0}$ in terms of gamma functions,
\be\label{pnfjgen}
\p^nf^j = \sum_{r=0}^n\frac{\Gamma(j+1)}{\Gamma(j-r+1)}\,f^{j-r}\,a_{n,r}\,.
\ee
Whenever $j\leq n$, the sum truncates at $r=j$ due to the poles in $\Gamma(j-r+1)$, reproducing \eqref{pnfj}. Repeating the steps in \eqref{finiteinv} using \eqref{pnfjgen} leads to
\be\label{goodboi}
\sum_{j=0}^k\frac{(-f)^{k-j}\p^nf^{j}}{j!(k-j)!} = 0\,,\qquad k>n\,,
\ee
for arbitrary smooth $f(z)$. This identity is essentially a consequence of the fact that there are no terms of $O(t^{n+1})$ or higher in \eqref{ansatz}.

\bibliographystyle{JHEP}
\bibliography{cope}

\end{document}